\documentclass{article}
\usepackage{iclr2027_conference,times}

\usepackage{amsmath,amsfonts,bm}

\def\eqref#1{equation~\ref{#1}}

\def\1{\bm{1}}

\DeclareMathAlphabet{\mathsfit}{\encodingdefault}{\sfdefault}{m}{sl}
\SetMathAlphabet{\mathsfit}{bold}{\encodingdefault}{\sfdefault}{bx}{n}

\usepackage{amsmath}
\usepackage{amssymb}
\usepackage{array}
\usepackage{booktabs}
\usepackage{multirow}
\usepackage{graphicx}
\usepackage{float}
\usepackage{placeins}
\usepackage{hyperref}
\usepackage{url}
\hypersetup{hidelinks}

\title{Recommendation World Models for Future-State Control}

\author{Jinfeng Xu$^{1}$, Zheyu Chen$^2$, Ziyue Peng$^3$, Jianheng Tang$^4$, Zheng Lin$^5$, Jing Yang$^6$, \\\textbf{Puzhen Wu}$^7$, \textbf{Zheng Xing}$^8$, \textbf{Victor C. M. Leung}$^1$\thanks{Corresponding Author.}\\ 
 \\
$^1$The University of British Columbia, Canada;
$^2$The Hong Kong Polytechnic University, Hong Kong; \\
$^3$The Hong Kong University of Science and Technology, Hong Kong;
$^4$Peking University, China; \\
$^5$University of Luxemburg, Luxemburg;
$^6$University Malaya, Malaysia; \\
$^7$The University of Hong Kong, Hong Kong;
$^8$Shenzhen University, China; 
}

\newcommand{\method}{UA-TWM}

\iclrfinalcopy
\begin{document}

\maketitle

\begin{abstract}
Sequential recommendation optimizes which items to rank, while each displayed slate also shapes subsequent feedback and user state. We study how a trained ranker can support decisions about these future consequences. We introduce \method{}, a utility-anchored world-model interface that constructs nearby slate actions, estimates their target-relevant consequences, and selects an alternative subject to utility constraints. The reference slate serves as a fallback when no alternative qualifies. A logged-replay instantiation combines utility and target-gain estimates with calibrated failure-risk prediction; a closed-loop instantiation uses one-step state-action prediction and updates its decisions after observed feedback. We evaluate transfer across twelve sequential backbones on MovieLens-25M and KuaiRand-Pure, and repeated target-directed interaction in KuaiSim. Attaching the interface improves Recall@20, NDCG@20, and future-state alignment for every matched logged backbone. Selection ablations reveal the utility and risk costs of aggressive target pursuit, while closed-loop diagnostics isolate the contribution of action-conditioned prediction. Local consequence modeling thus enables target-aware selection around a trained sequential ranker. 
\end{abstract}

\section{Introduction}

Sequential recommenders have become strong ranking models. Given a user history, they score the next item and return a slate with high short-term utility \citep{hidasi2016gru4rec,xu2025enhancing,sun2019bert4rec,zhou2020s3rec,tan2021sparse,xu2026CAMMSR}. Deployment adds a different question. A displayed slate changes the feedback available to the system and the user state used by the next recommendation step. Two slates can be close in Recall or NDCG while inducing different future category mixtures, repeat exposure, or interest concentration. A next-item score therefore does not, by itself, order the future consequences of candidate slate actions.

The difference is easiest to see at the slate level. Suppose two candidate slates both contain items that are plausible next clicks for the same user. One slate may concentrate the next session around a narrow genre, while the other keeps the same expected utility but broadens the user's subsequent consumption mixture. Choosing between these slates requires estimates of their future effects and the utility cost of steering toward a target. Item scores identify relevant candidates; evaluating their consequences requires conditioning on the slate as a whole and on the current user state.

This gap is central for recommendation systems that need to reason about future-state statistics rather than a single next click. A platform may want to preserve utility while reducing concentration, increasing exposure balance, or moving a user toward a requested content mixture. These targets are properties of trajectories produced by slate actions and feedback, not labels on isolated items. Interactive simulators and long-horizon recommendation methods make this feedback loop explicit \citep{ie2019recsim,xu2026survey,shi2019virtualtaobao,zhao2023kuaisim,gao2023cirs,gao2023dorl,xu2025mentor}, but most sequential rankers expose only a local ranking surface. A decision layer around a strong ranker can evaluate these consequences locally: it compares plausible slates by expected utility, target movement, and estimated failure risk.

List diversity and calibration provide useful ways to construct candidate slates. Consequence-aware selection adds a temporal question: which of these candidates is predicted to move the future statistic while preserving utility? The ranking model supplies a relevance-based reference, candidate generation exposes alternative actions, and consequence prediction evaluates their effects. This decomposition allows the decision layer to use the structure learned by a sequential ranker while optimizing an explicit target beyond next-item relevance.

We introduce \method{}, a utility-anchored recommendation world-model interface. A frozen recommender supplies the anchor slate, and a local consequence layer scores alternative slates against that reference. In logged replay, the interface combines utility and target-gain estimates with a calibrated failure-risk predictor. In closed-loop control, a learned state-action model predicts one-step consequences, and the controller selects a target-improving slate subject to a utility check. The anchor provides a fallback when no candidate passes selection. After a non-anchor selection, observed feedback determines whether the predicted target improvement and utility conditions hold.

We evaluate transfer and interaction in complementary settings. MovieLens-25M and KuaiRand-Pure provide chronological logs for matched comparisons across twelve sequential backbones. KuaiSim provides repeated interaction: the controller predicts one step ahead, executes a slate, observes feedback, and updates its state over ten steps. The logged comparisons measure alignment with historical future behavior; the simulator measures target movement following executed slates. Each simulated decision uses a one-step prediction within the ten-step interaction.

Our contributions are threefold. First, we formulate future-state-aware recommendation as selective slate-action control with anchor-relative utility and target criteria. Second, we instantiate this formulation as a local consequence interface that separates candidate generation, prediction, and constrained selection. Third, we evaluate both cross-backbone transfer and closed-loop behavior: matched logged comparisons improve Recall@20, NDCG@20, and future-state alignment across two public domains, while action-conditioning and selection ablations examine the mechanisms behind those decisions. This approach reuses the ranker's relevance estimates while learning how candidate actions affect the quantities needed for target-directed selection.

\section{Related Work}

\paragraph{Sequential recommendation as utility anchors.}
Sequential recommenders learn user-history representations for next-item ranking. Recurrent, convolutional, self-attentive, bidirectional, transition-based, self-supervised, multi-interest, diffusion, and flow-based models define heterogeneous ranking functions \citep{hidasi2016gru4rec,tang2018caser,kang2018sasrec,sun2019bert4rec,yuan2019nextitnet,rendle2010factorizing,he2017translation,liu2018stamp,fan2021lighter,zhou2020s3rec,tan2021sparse,chen2025adrec,shi2026fave}. These functions supply the relevance scores and reference slates for our local action search. Matched evaluations hold each backbone fixed and measure the effect of consequence-aware selection around it.

\paragraph{Controllable and interactive recommendation.}
Goal-conditioned recommendation and reinforcement-learning approaches make desired outcomes or long-horizon value explicit \citep{wang2021goalrec,li2024goalconditionedrec,gao2025mocdt,ie2019slateq,zhao2018deers,bai2019modelbasedrec,gao2023cirs,gao2023dorl}. Simulators and randomized recommendation logs provide complementary resources for studying user feedback \citep{ie2019recsim,shi2019virtualtaobao,gao2022kuairand,zhao2023kuaisim}. Our controller searches candidate slates near the ranker's reference recommendation, evaluates their target movement and utility cost, and retains that reference when the selection criteria are unmet.

\paragraph{World models for selective action.}
World models learn dynamics for prediction, planning, or control \citep{sutton1991dyna,ha2018worldmodels,chua2018pets,janner2019mbpo,hafner2020dreamer,hafner2021dreamerv2,schrittwieser2020muzero}. Return-conditioned sequence modeling provides another route to decision making \citep{chen2021decisiontransformer}. In recommendation, the state summarizes user history, the action is a slate, and the predicted quantities describe subsequent feedback and consumption. Recent LLM-based user simulators and judges address related evaluation problems \citep{bonin2025llmjudge,bougie2026alignuser}. Our closed-loop instantiation uses one-step state-action prediction for repeated constrained selection. Appendix Section~\ref{app:closest-work} expands the comparison with related recommendation approaches.

\section{Methodology}
\label{sec:methodology}

\subsection{Decision Problem and Evidence Regimes}

We model repeated recommendation as a controlled process. At time $t$, the system summarizes the history $h_t=(a_0,y_0,\ldots,a_{t-1},y_{t-1})$ into a state $s_t=g_\psi(h_t)$, selects a slate $a_t=(i_{t,1},\ldots,i_{t,K})$, and observes feedback $y_t$. The environment induces
\begin{equation}
  P^\star(s_{t+1},y_t\mid s_t,a_t)
  =
  P^\star(y_t\mid s_t,a_t)
  P^\star(s_{t+1}\mid s_t,a_t,y_t).
  \label{eq:true-kernel}
\end{equation}
A target map $\phi$ extracts a horizon-$H$ future-state statistic from the induced trajectory, such as a category mixture, concentration score, or repeat exposure statistic. For a target $z^\star$, the interventional target loss and horizon utility of a policy $\pi$ are
\begin{align}
  L_H(\pi;s,z^\star)
  &=
  \mathbb{E}_{P^\star,\pi}
  \left[D\!\left(\phi(\tau_{t:t+H}),z^\star\right)\right],
  &
  V_H(\pi;s,z^\star)
  &=
  \mathbb{E}_{P^\star,\pi}
  \left[\sum_{h=0}^{H-1} r(y_{t+h},a_{t+h})\right].
  \label{eq:target-utility}
\end{align}
The utility-safe control problem is to reduce $L_H$ without dropping below a strong recommendation anchor $\pi_{\mathrm{anc}}$:
\begin{equation}
  \min_{\pi}
  \mathbb{E}_{(s,z^\star)}
  [L_H(\pi;s,z^\star)]
  \quad
  \mathrm{s.t.}\quad
  \mathbb{E}_{(s,z^\star)}
  [V_H(\pi;s,z^\star)]
  \geq
  \mathbb{E}_{(s,z^\star)}
  [V_H(\pi_{\mathrm{anc}};s,z^\star)]-\epsilon_u .
  \label{eq:population-objective}
\end{equation}

We approximate this control problem through local action-conditioned prediction. Given the current state, a target, and candidate slates around the anchor, the consequence model estimates target movement and the selector checks utility preservation. The model scores the candidate slates available to the controller. The policy selects a qualifying alternative or returns the anchor.

Closed-loop rollout evaluates the induced trajectory under the selected policy. Logged replay evaluates a retrospective proxy: prefix-generated candidate slates are compared with held-out future positives collected under the historical policy. The former measures action-induced target movement within the simulator; the latter measures future-statistic alignment under chronological replay. Logged positives do not identify the transition distribution of an unshown slate.

Both regimes evaluate utility and target conditions relative to the same anchor. In rollout, failure concerns induced utility and trajectory statistics. In replay, failure concerns ranking utility and alignment with the historical future-positive mixture. Coverage records non-anchor selection frequency, and invalidity records failures among those selections.

\subsection{\method{}: A Utility-Anchored World-Model Interface}

\method{} converts a trained sequential recommender into a selective slate-action controller. The backbone is frozen and supplies the base slate $A^0_{u,t}$. The consequence layer is trained only around candidate slates near this anchor. It predicts utility gain, future-state gain, and support risk, then uses these predictions to decide whether to select a non-anchor slate or return the anchor.

\begin{figure}[!t]
  \centering
  \includegraphics[width=\linewidth]{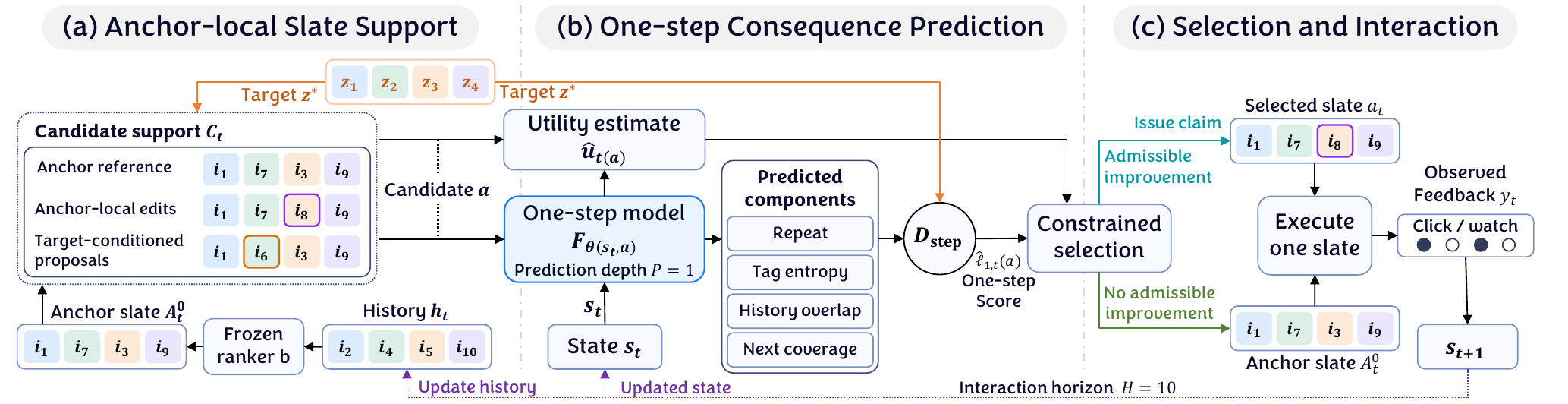}
  \caption{Closed-loop \method{}: (a) anchor-local slate proposals, (b) one-step consequence prediction, and (c) constrained selection with feedback. Prediction depth is $P=1$; interaction lasts $H=10$ steps. Slates are schematic.}
  \label{fig:method-schematic}
  \vskip -0.1in
\end{figure}

\paragraph{Candidate support.}
Let $c_i\in\{0,1\}^{d}$ denote the observable multi-hot genre or tag vector of item $i$, and let $\mathcal{P}_M(h)$ be the top-$M$ pool returned by the frozen backbone after masking items already seen in the prefix. Profiles are normalized after summing item vectors. The decision-time target is $z^\star$ in a controlled rollout and $z^\dagger_{u,t}$ in logged replay. The logged target is prefix-only:
\begin{equation}
  z^\dagger_{u,t}
  =
  \frac{
  \sum_{i\in \mathcal{H}^{+,\mathrm{train}}_{u,t}} c_i
  +\omega
  \sum_{i\in \mathcal{H}^{+,\mathrm{calib}}_{u,t}} c_i
  }{
  \left\|
  \sum_{i\in \mathcal{H}^{+,\mathrm{train}}_{u,t}} c_i
  +\omega
  \sum_{i\in \mathcal{H}^{+,\mathrm{calib}}_{u,t}} c_i
  \right\|_1
  },
  \label{eq:prefix-target}
\end{equation}
with $\omega=2$ weighting the more recent calibration block. Calibration scoring uses training history; test scoring uses the completed training and calibration blocks. An empty profile returns the zero vector. Appendix~\ref{app:logged-temporal-protocol} specifies the decision boundaries, normalization, and future windows. The steering score used for target-conditioned proposals is
\begin{equation}
  q(i,z,h)=\langle c_i,z\rangle ,
  \label{eq:steering-score}
\end{equation}
so the target signal can change the order of plausible backbone candidates without introducing items outside the anchor's local support. Logged candidates are generated by
\begin{equation}
  A^\beta(h,z)
  =
  \operatorname{TopK}_{i\in\mathcal{P}_M(h)}
  \left[
    \tilde s_b(i\mid h)+\beta q(i,z,h)
  \right],
  \qquad
  \beta\in\mathcal{B}.
  \label{eq:wm-candidate-grid}
\end{equation}
Here $\tilde s_b$ is the within-pool standardized backbone score, and $\beta$ is selected on calibration users. The closed-loop instantiation constructs candidates from two proposal sources. Anchor-local edits keep most items from $A^0$ and replace a small number of low-contribution items with high-utility candidates whose item/tag summaries improve the requested statistic. Graph-semantic replacements are fixed proposal generators over observable state and item metadata. They form candidate pools from immediate click scores, history-repeat penalties, history-tag similarity penalties, and within-slate diversity penalties; then they instantiate repeat-oriented, balanced, and diversity-oriented slates under preset click-floor variants. Both proposal sources use the state and item metadata available before execution; the consequence model then scores the resulting slates. In the KuaiSim diagnostic, each decision has nine candidate actions: the anchor, two anchor-local edits, and six graph-semantic replacement slates. Candidate construction defines the controller's local search space around the utility anchor.

\paragraph{Replay consequence heads.}
Let $\mathcal{F}^{+}_{u,t}$ contain the positive events in the held-out block following the decision boundary. The logged future statistic is
\begin{equation}
  \tilde z_{u,t}=\phi(\mathcal{F}^{+}_{u,t})
  \label{eq:logged-target-state}
\end{equation}
and uses the full variable-length block to supply calibration labels or test outcomes. The $H$-step horizon applies to closed-loop interaction. A candidate slate $A$ is scored by
\begin{equation}
  U(A,u)=\operatorname{Recall@20}(A,\mathcal{F}^{+}_{u,t}),
  \qquad
  S(A,u)=\|\phi(A)-\tilde z_{u,t}\|_1 .
  \label{eq:logged-utility-state-loss}
\end{equation}
The consequence labels compare each candidate with the anchor:
\begin{align}
  \Delta_U(u,A)&=U(A,u)-U(A^0,u),&
  \Delta_Z(u,A)&=S(A^0,u)-S(A,u).
  \label{eq:realized-gains}
\end{align}
The logged implementation estimates target and utility margins from prefix statistics and backbone scores. The predicted state movement is the prefix-target margin
\begin{equation}
  \hat g_Z(u,A)
  =
  \|\phi(A^0)-z^\dagger_{u,t}\|_1-\|\phi(A)-z^\dagger_{u,t}\|_1 ,
  \label{eq:logged-state-head}
\end{equation}
and the predicted utility margin is the mean standardized backbone-score margin between $A$ and $A^0$:
\begin{equation}
  \hat g_U(u,A)
  =
  \frac{1}{K}\sum_{i\in A}\tilde s_b(i\mid h_t)
  -
  \frac{1}{K}\sum_{i\in A^0}\tilde s_b(i\mid h_t).
  \label{eq:logged-utility-head}
\end{equation}
The risk head is learned from calibration candidates. Its feature vector contains only prefix- and candidate-support quantities: $\beta$, $\hat g_Z$, $\hat g_U$, target concentration $\max_j z^\dagger_{u,t,j}$, $\hat g_Z\max_j z^\dagger_{u,t,j}$, $\beta\hat g_Z$, $\hat g_U\max_j z^\dagger_{u,t,j}$, $\max(0,-\hat g_U)$, $\hat g_Z^2$, and $\hat g_U^2$. The held-out statistic $\tilde z_{u,t}$ is not part of $x(u,A)$. The consequence interface therefore has three action-conditioned outputs,
\begin{equation}
  \hat g_U(u,A),\qquad
  \hat g_Z(u,A),\qquad
  \hat r(u,A),
  \label{eq:consequence-heads}
\end{equation}
where $\hat r$ predicts failure of either the target-gain margin or the per-user utility floor:
\begin{equation}
  y_{\mathrm{risk}}(u,A)
  =
  \mathbf{1}
  [\Delta_Z(u,A)<\delta_Z\ \lor\ \Delta_U(u,A)<-\epsilon_{\mathrm{user}}].
  \label{eq:logged-risk-label}
\end{equation}
The reported logged runs train $\hat r$ as an $L_2$-regularized logistic classifier with binary cross entropy,
\begin{equation}
  \mathcal{L}_{\mathrm{risk}}
  =
  \operatorname{BCE}(\hat r,y_{\mathrm{risk}})
  +\lambda\|\theta_r\|_2^2 .
  \label{eq:risk-loss}
\end{equation}
For the split-risk diagnostic, the binary label is decomposed into $y_U=\mathbf{1}[\Delta_U<-\epsilon_{\mathrm{user}}]$ and $y_Z=\mathbf{1}[\Delta_Z<\delta_Z]$, yielding two probabilities $\hat r_U$ and $\hat r_Z$ whose maximum defines the gate risk. Thresholds are chosen on calibration users. Test future positives never enter candidate generation or threshold selection.

\paragraph{Selective planning and claims.}
Let $\mathcal{C}_{\mathrm{adm}}(u)=\{A\in\mathcal{C}(u):\hat g_U(u,A)\geq\tau_U,\;\hat g_Z(u,A)\geq\tau_Z,\;\hat r(u,A)\leq\rho\}$ be the candidates that pass all three gates. At test time the logged selector returns
\begin{equation}
  \pi_{\mathrm{UA}}(u)
  =
  \begin{cases}
  \arg\max_{A\in\mathcal{C}_{\mathrm{adm}}(u)} \hat g_Z(u,A),
  &
  \mathcal{C}_{\mathrm{adm}}(u)\neq\varnothing,\\
  A^0_{u,t},&\text{if no candidate satisfies the gates}.
  \end{cases}
  \label{eq:logged-selector}
\end{equation}
For closed-loop evaluation, a non-anchor selection constitutes a target-control claim made before feedback is observed; anchor retention issues no claim. For a selected non-anchor action $a$ at state $s$ and target $z^\star$, the claim is valid when the induced rollout improves the target statistic while preserving the anchor utility:
\begin{equation}
  \Gamma^\star(s,z^\star,a)
  =
  \mathbf{1}
  \left[
    L_H(a;s,z^\star)
    \leq
    L_H(a_{\mathrm{anc}};s,z^\star)-\delta_z
    \land
    V_H(a;s,z^\star)
    \geq
    V_H(a_{\mathrm{anc}};s,z^\star)-\epsilon_u
  \right].
  \label{eq:claim-event}
\end{equation}
Coverage is the fraction of evaluated scenarios with an issued non-anchor claim. Invalidity is the fraction of issued claims that fail Equation~\ref{eq:claim-event}; unsuccessful claims remain in the denominator. Returning the anchor contributes to neither the claim count nor the invalid-claim count.

\paragraph{Calibration and operating points.}
The thresholds $(\tau_U,\tau_Z,\rho)$ define the logged selector's operating point and are chosen on calibration users before test evaluation. The utility threshold limits the predicted utility cost, the target threshold sets the required predicted movement, and the risk threshold filters candidates by estimated failure. For fixed utility and target thresholds, a tighter risk gate retains fewer candidates. Selection maximizes predicted target gain over those that remain.

The ablations isolate these decisions. Removing the utility floor tests the cost of target movement; removing the state head tests the contribution of target prediction; removing the risk gate tests whether gain estimates suffice to select reliable candidates. Removing fallback forces a non-anchor choice when the full selector would abstain, exposing the cost of intervention on those cases.

\paragraph{One-step prediction and repeated interaction.}
The closed-loop controller fits a ridge state-action model to offline tuples of aggregate states, slate summaries, feedback, and next-state statistics:
\begin{equation}
  \min_\theta
  \sum_{(s,a,y,s')\in\mathcal{D}_{\mathrm{dyn}}}
  \left\|
  F_\theta(s,a)
  -
  \psi(y,s')
  \right\|_2^2 +\lambda_{\mathrm{dyn}}\|\theta\|_2^2,
  \label{eq:dynamics-training}
\end{equation}
Here $\lambda_{\mathrm{dyn}}$ is the ridge regularization coefficient. The prediction depth is $P=1$ and the interaction horizon is $H=10$. A readout $q_1$ extracts predicted repeat exposure, tag entropy, history-tag overlap, and next-state coverage. The one-step target score and anchor-relative gain are
\begin{align}
  \hat z^{(1)}_t(a) &= q_1\!\left(F_\theta(s_t,a)\right), \qquad
  \hat\ell_{1,t}(a) = D_{\mathrm{step}}\!\left(\hat z^{(1)}_t(a),z^\star\right), \\
  \hat g^{(1)}_{Z,t}(a) &= \hat\ell_{1,t}(a_t^0)-\hat\ell_{1,t}(a),
  \label{eq:rollout-prediction}
\end{align}
where $D_{\mathrm{step}}$ compares predicted components with the target. The controller minimizes $\hat\ell_{1,t}$ among admissible candidates, using an immediate click estimate for the utility check, and retains the anchor unless its target score improves. After executing one slate, observed feedback updates the state and candidates are regenerated. This repeats for ten environment steps. The trajectory-level $L_H$ and $V_H$ evaluate the resulting interaction; selection uses one-step predictions throughout.

\section{Experiments}
\label{sec:experiments}

\subsection{Evaluation Protocol}

The experiments examine three questions: does the consequence interface improve recommendations across backbones, which selection components account for the improvement, and does action-conditioned prediction support target-directed interaction? We use chronological replay for the cross-backbone comparison and KuaiSim rollouts for the interaction test. Alternatives and ablations share the candidate support of the corresponding full method.

MovieLens-25M uses a frozen chronological 70/10/20 split over 162,541 users and 25M ratings. Utility is Recall@20 and NDCG@20 over future high-rated items. Future-state loss is the $L_1$ distance between the genre mixture of the deployed slate and the genre mixture of future high-rated items. KuaiRand-Pure uses a chronological short-video split over 17,269 users, 7,583 videos, and 111 tag-state categories. A positive interaction denotes click or long view. Future-state loss is the $L_1$ distance between the deployed slate's tag mixture and the future positive-video tag mixture. Within each dataset, all methods share the same histories, item vocabulary, top-200 candidate pool, calibration split, slate size, and target-state definition. Logged replay uses $K=20$, maximum history length 100, candidate weights $\beta\in\{0.05,0.1,0.2,0.3,0.5,0.75,1.0\}$, and calibration-selected utility, state, and risk gates. Closed-loop rollout uses prediction depth $P=1$, interaction horizon $H=10$, and slate size 6.

At each logged decision boundary, the target profile uses the available positive history, and the selected slate is evaluated against the following held-out positive block. This block has variable length across users. The profile weights each calibration-block event by $\omega=2$ at test time; calibration scoring uses the training block alone. Item tags are summed before normalization. Appendix~\ref{app:logged-temporal-protocol} gives the event-level construction.

The logged protocol evaluates twelve sequential recommendation anchors grouped by model family. The canonical sequence encoders are GRU4Rec, SASRec, BERT4Rec, and NextItNet, which represent recurrent, self-attentive, bidirectional, and convolutional next-item modeling \citep{hidasi2016gru4rec,kang2018sasrec,sun2019bert4rec,yuan2019nextitnet}. The second group covers structurally distinct ranking surfaces: FPMC and TransRec supply Markov and metric-transition anchors, STAMP supplies a session-attention anchor, LightSANs and S3Rec supply attention and self-supervised sequence anchors, and SINE supplies a multi-interest anchor \citep{rendle2010factorizing,he2017translation,liu2018stamp,fan2021lighter,zhou2020s3rec,tan2021sparse}. The third group adds recent generative sequential recommenders: ADRec uses diffusion-style sequence prediction and FAVE uses flow-based average-velocity prediction \citep{chen2025adrec,shi2026fave}. For external repositories, we convert the same frozen histories into each required training format and evaluate ranked slates with the same replay evaluator used for all anchors. For every backbone, Base is the trained recommender and +UA-TWM is the same recommender after attaching the consequence layer and constrained selector. The matched with/without design isolates the decision interface because the backbone parameters are unchanged.

We report ranking utility, future-state loss, coverage, and invalidity. Recall@20 and NDCG@20 measure retrieval and rank-sensitive utility, respectively. Future-state loss measures discrepancy from the regime-specific target statistic. Logged coverage counts non-anchor selections over evaluated users; closed-loop coverage counts issued target-scenario claims. Invalidity is conditional on issuing a claim. We interpret coverage together with invalidity because anchor retention can reduce failures without improving target control.

\subsection{Logged Transfer Across Backbones}

Table~\ref{tab:main-logged-results} reports the matched logged test. Across two recommendation domains and twelve backbone families, attaching \method{} improves Recall@20, NDCG@20, and future-state alignment for every matched backbone. On MovieLens-25M, Recall@20 gains range from 0.0119 to 0.0144, NDCG@20 gains range from 0.0028 to 0.0102, and future-state $L_1$ drops by 0.037 to 0.046. On KuaiRand-Pure, Recall@20 gains range from 0.0054 to 0.0068, NDCG@20 gains range from 0.0018 to 0.0037, and future-state loss drops by 0.064 to 0.082. In relative terms, the median improvements are 5.9\% Recall@20 and 5.9\% NDCG@20 on MovieLens-25M, and 6.4\% Recall@20 and 6.5\% NDCG@20 on KuaiRand-Pure; the mean relative gains are larger because ADRec starts from a lower MovieLens utility point. The improvement spans classical Markov models, session and attention models, self-supervised and multi-interest models, and diffusion and flow-based anchors.

\begin{table}[!t]
  \caption{Logged transfer across sequential recommendation backbones. Attaching \method{} improves ranking utility and future-state alignment for every matched backbone on both datasets. R@20 is Recall@20, N@20 is NDCG@20, and $S$ is future-state $L_1$ loss. I/C reports replay-unsafe rate / coverage in percent for non-anchor \method{} claims.}
  \label{tab:main-logged-results}
  \centering
  \vskip 0pt
  \small
  \setlength{\tabcolsep}{0pt}
  \renewcommand{\arraystretch}{1.02}
  \begin{tabular*}{\linewidth}{@{\extracolsep{\fill}}llrrrrrrrr@{}}
    \toprule
    & & \multicolumn{4}{c}{MovieLens-25M}
      & \multicolumn{4}{c}{KuaiRand-Pure} \\
    \cmidrule(lr){3-6}\cmidrule(lr){7-10}
    Backbone & Policy & R@20 $\uparrow$ & N@20 $\uparrow$ & $S\downarrow$ & I/C
      & R@20 $\uparrow$ & N@20 $\uparrow$ & $S\downarrow$ & I/C \\
    \midrule
    \multicolumn{10}{@{}l}{\textit{Canonical sequence encoders}} \\
    GRU4Rec & Base & 0.1177 & 0.0338 & 0.881 & -- & 0.0645 & 0.0263 & 1.461 & -- \\
     & +UA-TWM & \textbf{0.1304} & \textbf{0.0375} & \textbf{0.840} & 3.4/56.2 & \textbf{0.0703} & \textbf{0.0287} & \textbf{1.389} & 5.1/95.7 \\
    \addlinespace[0.4mm]
    SASRec & Base & 0.1067 & 0.0371 & 0.886 & -- & 0.0656 & 0.0257 & 1.453 & -- \\
     & +UA-TWM & \textbf{0.1194} & \textbf{0.0415} & \textbf{0.845} & 3.6/54.7 & \textbf{0.0714} & \textbf{0.0280} & \textbf{1.381} & 5.2/96.0 \\
    \addlinespace[0.4mm]
    BERT4Rec & Base & 0.2454 & 0.0989 & 0.749 & -- & 0.1080 & 0.0389 & 1.392 & -- \\
     & +UA-TWM & \textbf{0.2581} & \textbf{0.1040} & \textbf{0.708} & 5.1/94.2 & \textbf{0.1138} & \textbf{0.0410} & \textbf{1.320} & 6.0/98.2 \\
    \addlinespace[0.4mm]
    NextItNet & Base & 0.1194 & 0.0382 & 0.878 & -- & 0.0652 & 0.0256 & 1.455 & -- \\
     & +UA-TWM & \textbf{0.1321} & \textbf{0.0423} & \textbf{0.837} & 3.5/55.8 & \textbf{0.0710} & \textbf{0.0279} & \textbf{1.383} & 5.0/95.5 \\
    \addlinespace[0.6mm]
    \multicolumn{10}{@{}l}{\textit{Transition, session, attention, and multi-interest anchors}} \\
    FPMC & Base & 0.2184 & 0.0534 & 0.757 & -- & 0.1036 & 0.0352 & 1.398 & -- \\
     & +UA-TWM & \textbf{0.2328} & \textbf{0.0569} & \textbf{0.711} & 3.5/57.8 & \textbf{0.1104} & \textbf{0.0375} & \textbf{1.319} & 5.0/98.9 \\
    \addlinespace[0.4mm]
    TransRec & Base & 0.2127 & 0.0451 & 0.758 & -- & 0.0925 & 0.0311 & 1.426 & -- \\
     & +UA-TWM & \textbf{0.2257} & \textbf{0.0479} & \textbf{0.716} & 3.8/53.1 & \textbf{0.0983} & \textbf{0.0331} & \textbf{1.344} & 5.5/95.2 \\
    \addlinespace[0.4mm]
    STAMP & Base & 0.2216 & 0.0549 & 0.755 & -- & 0.1024 & 0.0354 & 1.397 & -- \\
     & +UA-TWM & \textbf{0.2341} & \textbf{0.0580} & \textbf{0.718} & 4.1/52.4 & \textbf{0.1081} & \textbf{0.0374} & \textbf{1.333} & 5.9/96.1 \\
    \addlinespace[0.4mm]
    LightSANs & Base & 0.2441 & 0.0959 & 0.749 & -- & 0.1029 & 0.0364 & 1.396 & -- \\
     & +UA-TWM & \textbf{0.2560} & \textbf{0.1006} & \textbf{0.709} & 5.2/93.8 & \textbf{0.1084} & \textbf{0.0383} & \textbf{1.322} & 6.1/97.4 \\
    \addlinespace[0.4mm]
    S3Rec & Base & 0.2492 & 0.0952 & 0.749 & -- & 0.1148 & 0.0392 & 1.368 & -- \\
     & +UA-TWM & \textbf{0.2612} & \textbf{0.0998} & \textbf{0.708} & 5.1/94.4 & \textbf{0.1202} & \textbf{0.0410} & \textbf{1.304} & 6.0/98.6 \\
    \addlinespace[0.4mm]
    SINE & Base & 0.2199 & 0.0556 & 0.757 & -- & 0.1045 & 0.0358 & 1.397 & -- \\
     & +UA-TWM & \textbf{0.2320} & \textbf{0.0587} & \textbf{0.719} & 3.3/63.5 & \textbf{0.1100} & \textbf{0.0377} & \textbf{1.329} & 5.6/99.0 \\
    \addlinespace[0.6mm]
    \multicolumn{10}{@{}l}{\textit{Recent generative sequence backbones}} \\
    ADRec & Base & 0.0245 & 0.0191 & 0.819 & -- & 0.0591 & 0.0368 & 1.430 & -- \\
     & +UA-TWM & \textbf{0.0372} & \textbf{0.0290} & \textbf{0.778} & 4.2/41.5 & \textbf{0.0649} & \textbf{0.0404} & \textbf{1.358} & 5.8/94.8 \\
    \addlinespace[0.4mm]
    FAVE & Base & 0.2528 & 0.2026 & 0.753 & -- & 0.0483 & 0.0311 & 1.448 & -- \\
     & +UA-TWM & \textbf{0.2655} & \textbf{0.2128} & \textbf{0.712} & 5.0/94.0 & \textbf{0.0541} & \textbf{0.0348} & \textbf{1.376} & 6.1/98.1 \\
    \bottomrule
  \end{tabular*}
  \vskip -0.1in
\end{table}

The gains persist on the strongest anchors. On MovieLens-25M, S3Rec improves from 0.2492 to 0.2612 Recall@20 and from 0.0952 to 0.0998 NDCG@20; FAVE improves from 0.2528 to 0.2655 Recall@20. On KuaiRand-Pure, S3Rec improves from 0.1148 to 0.1202 Recall@20 while state loss decreases from 1.368 to 1.304. The matched gains on S3Rec and FAVE show that consequence-aware selection remains useful around high-utility ranking anchors.

Selection frequency differs substantially across the two domains. MovieLens coverage ranges from 41.5\% to 94.4\%, with replay-unsafe rates of 3.3\%--5.2\%. KuaiRand coverage ranges from 94.8\% to 99.0\%, with unsafe rates of 5.0\%--6.1\%. KuaiRand's gains accompany near-universal non-anchor selection, while MovieLens retains the anchor more often. The same-support comparisons below examine how the selector trades target gain against utility and failure risk.

\subsection{Ranking and Future-State Ordering}

\begin{figure}[!t]
  \centering
  \includegraphics[width=0.98\linewidth]{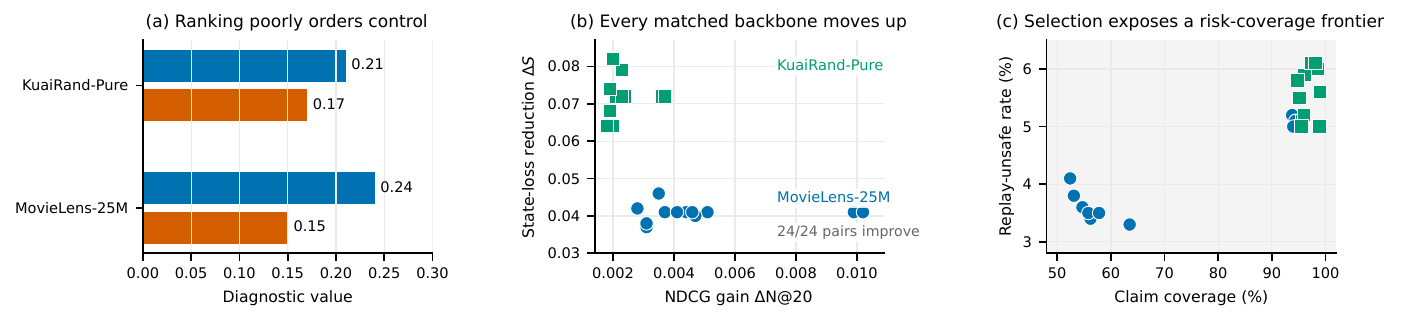}
  \caption{Empirical observation behind utility-anchored recommendation world models. Panel (a) measures the mismatch between ranking order and future-state order. Panel (b) shows matched NDCG@20 and future-state improvements for the twenty-four logged dataset-backbone pairs. Panel (c) reports replay-unsafe rate and coverage for selected non-anchor claims.}
  \label{fig:empirical-observation}
  \vskip -0.1in
\end{figure}

For each backbone, we evaluate the base slate both as a ranking decision and as a future-state action. The diagnostic compares the utility ordering induced by Recall@20/NDCG@20 with the control ordering induced by future-state loss and replay-unsafe rate. Low agreement means that a high-utility slate can still be a poor action for the requested future statistic. Figure~\ref{fig:empirical-observation} summarizes this mismatch and the matched logged gains after adding the consequence layer.

The ordering mismatch concerns the choice among plausible actions. Relevance determines which items enter the candidate pool, but it does not fully determine which slate best matches the future statistic. A separate consequence score can therefore change the selected action even when the candidate items and backbone remain fixed. Panel (b) connects this observation to the matched experiment: the twenty-four dataset-backbone pairs improve in both NDCG and state alignment. Panel (c) then exposes the remaining selection failures through the coverage--unsafe-rate relationship.

\subsection{Alternatives, Ablations, and Model Diagnostics}

\begin{figure}[!t]
  \centering
  \includegraphics[width=0.78\linewidth]{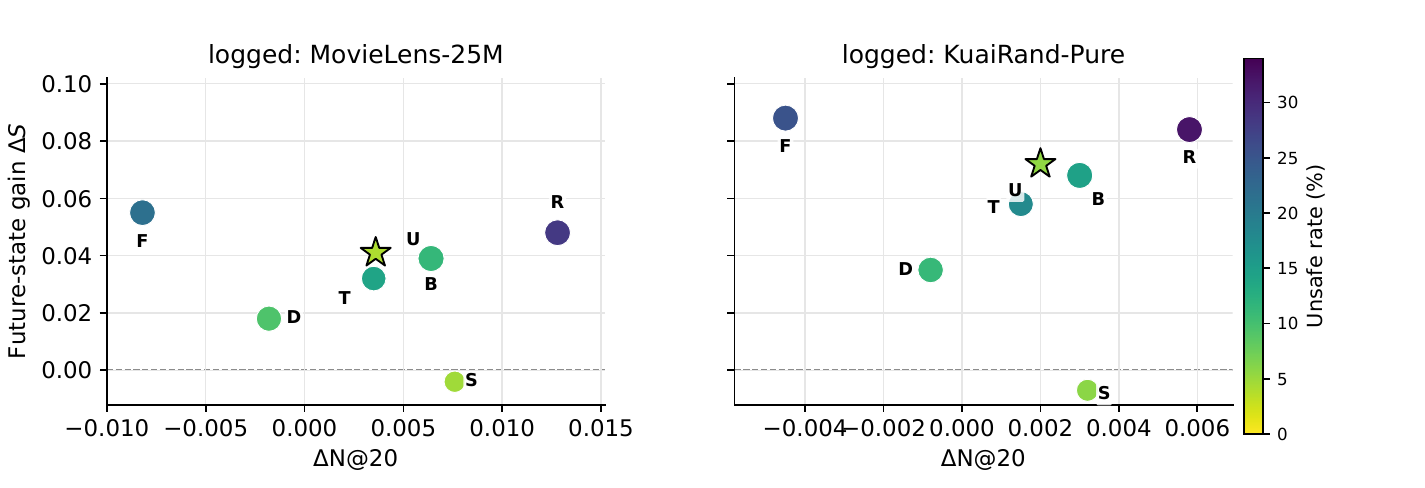}
  \caption{Logged replay alternatives and ablations. T: target-only reranking; D: diversity reranking; U: full \method{}; F: without the utility floor; S: without the state head; R: without the risk gate; B: without fallback. The star denotes \method{}, marker size encodes coverage, and color encodes replay-unsafe rate.}
  \label{fig:logged-frontier}
  \vskip -0.1in
\end{figure}

We compare the full selector with direct target-conditioned reranking, generic diversity reranking, scalar scoring, and ablations of the utility floor, state head, risk gate, and fallback. These comparisons share the histories, candidate pool, calibration split, and held-out future positives of Table~\ref{tab:main-logged-results}. Figure~\ref{fig:logged-frontier} separates the effects: removing the utility floor favors larger state movement at a ranking cost; removing the state head reduces target improvement; and removing the risk gate admits more unsafe claims. The full selector balances the three quantities by ranking candidates within a constrained feasible set.

The cross-backbone comparison in Appendix Table~\ref{tab:same-support-ablation} makes this trade-off concrete. On MovieLens FPMC, aggressive selection reduces state loss by 0.051 with 12.5\% invalidity, whereas constrained selection reduces it by 0.046 with 3.5\% invalidity. On KuaiRand ADRec, the aggressive policy has negative Recall and NDCG changes; the constrained policy recovers positive changes in both while lowering invalidity from 10.5\% to 5.8\%. The constrained policy trades some target movement for lower invalidity and positive ranking gains.

Risk calibration provides a complementary diagnostic (Appendix Table~\ref{tab:wm-quality}). Split-risk gating lowers selected unsafe rates from 5.9\% to 4.1\% for MovieLens S3Rec and from 6.4\% to 4.8\% for KuaiRand LightSANs. The corresponding ECE changes are 0.135 to 0.099 and 0.098 to 0.097. Sensitivity experiments on these two settings retain positive ranking and state gains across $M\in\{100,200,300\}$, reduced proposal grids, and $\omega\in\{1,2,3\}$ (Appendix~\ref{app:logged-sensitivity}). With the policy fixed, shorter future windows increase invalidity; Figures~\ref{fig:logged-sensitivity}--\ref{fig:logged-window-sensitivity} report the pointwise trade-offs.

The fixed-policy window test further examines the time span represented by the logged target. On a common cohort with at least 20 test-positive events, the same selected slate is scored against the first 10 events, the first 20, or the full positive block. MovieLens state-loss reduction increases from 0.035 to 0.039 to 0.042, while unsafe rates fall from 8.8\% to 6.5\% to 4.7\%. KuaiRand follows the same direction. Coverage stays fixed because the decision prefix and policy are unchanged. The selected slates therefore align more closely with the longer observed consumption mixture in these cohorts.

\subsection{Closed-Loop Target-Control}

We evaluate closed-loop control on seven KuaiSim target scenarios (Appendix Table~\ref{tab:kuaisim-target-grid}), with ten seeds, prediction depth $P=1$, interaction horizon $H=10$, slate size 6, and a diversity-rerank anchor. Ridge dynamics are trained on offline transitions from separate behavior policies, leaving each evaluation seed out. At each step, the model scores the anchor, two local edits, and six graph-semantic replacements. KuaiSim executes the selected slate; observed feedback updates the state before candidates are regenerated.

Table~\ref{tab:closed-loop-diagnostic} compares the anchor, two target-aggressive semantic controllers, a utility-constrained reachable-frontier support selector (RFSS-U), a state-only dynamics ablation, and \method{}. RFSS-U is selected from a fixed support-frontier manifest. The state-only ablation holds the gates fixed and removes slate-action summaries from the learned predictor. Holding candidate support fixed tests whether action-conditioned prediction identifies useful alternatives under the same gates.

\method{} obtains click 42.936 and target $L_1$ 2.581. It makes four valid target-control claims, falls back on three unsupported or utility-unsafe targets, and makes zero invalid claims for every seed. Before fallback, the state-action controller improves target loss over the anchor by 0.374 with 95\% CI [0.251, 0.505]. The state-only ablation makes no claims and has a smaller gain of 0.201; the paired state-action advantage is 0.173 target $L_1$ with 95\% CI [0.044, 0.313].

\begin{table}[!t]
  \caption{Closed-loop control on KuaiSim under a shared target grid and evaluation protocol. Click and target loss are means; claims, invalid claims, and fallbacks are target-scenario counts.}
  \label{tab:closed-loop-diagnostic}
  \centering
  \vskip 0pt
  \footnotesize
  \setlength{\tabcolsep}{3.2pt}
  \renewcommand{\arraystretch}{1.01}
  \begin{tabular}{@{}llrrrrr@{}}
    \toprule
    Method & Role & Click & Target $L_1$ & Claims & Invalid & Fallback \\
    \midrule
    diversity reranker & high-click anchor & 42.892 & 2.819 & -- & -- & -- \\
    semantic diversity & target-aggressive control & 42.689 & 0.657 & -- & -- & -- \\
    semantic repeat & target-aggressive control & 42.725 & 0.429 & -- & -- & -- \\
    RFSS-U & support-frontier control reference & 42.789 & \textbf{0.417} & 5 & \textbf{0} & 2 \\
    state-only dynamics & selective ablation & 42.892 & 2.819 & 0 & \textbf{0} & 7 \\
    \method{} & state-action consequence interface & \textbf{42.936} & 2.581 & \textbf{4} & \textbf{0} & 3 \\
    \bottomrule
  \end{tabular}
  \vskip -0.1in
\end{table}

The operating points in Table~\ref{tab:closed-loop-diagnostic} reveal the role of the utility criterion. Semantic controls and RFSS-U achieve lower target loss with lower click utility than the anchor. \method{} retains click utility while making a smaller target improvement. The state-only ablation returns the anchor throughout, whereas action-conditioned scoring selects qualifying alternatives on four target scenarios.

Action summaries also improve ridge transition prediction: normalized MAE decreases from 0.491 to 0.266 on held-out seeds and from 0.433 to 0.345 on held-out policies (Appendix Table~\ref{tab:dynamics-prediction}). Random forests instead worsen on held-out policies, so the benefit depends on the predictor and transfer setting. The ridge predictor improves both held-out prediction and the target-gain comparison with the state-only controller. Appendix~\ref{app:closed-loop-details} reports target-family outcomes and decision traces.

\section{Conclusion}

\method{} connects sequential ranking to target-directed decisions by predicting the consequences of nearby slate actions and selecting them under utility constraints. Matched logged evaluations across twelve backbones and two domains improve ranking utility and future-state alignment. In KuaiSim, action-conditioned prediction supports repeated target-directed selection, with the anchor retained when no candidate qualifies.

\clearpage

\bibliography{iclr2027_conference}
\bibliographystyle{iclr2027_conference}

\appendix
\raggedbottom
\setlength{\textfloatsep}{12pt plus 2pt minus 2pt}
\setlength{\floatsep}{10pt plus 2pt minus 2pt}
\setlength{\intextsep}{10pt plus 2pt minus 2pt}
\section{Additional Details}
\label{app:additional-details}

The appendix is organized by evidence role. Appendix~\ref{app:evidence-map} states how each experimental setting supports the claims in the main text. Appendix~\ref{app:additional-logged-analyses} expands the logged replay diagnostics. Appendix~\ref{app:closest-work} separates \method{} from the closest simulator, reinforcement-learning, causal, goal-conditioned, and LLM-based recommendation lines. Appendix~\ref{app:closed-loop-details} gives the closed-loop mechanism details behind the KuaiSim diagnostic.

\subsection*{AI use statement}

We used generative AI tools for literature organization, manuscript editing, code navigation, and experiment-log summarization. The authors checked research claims, reported numbers, citation choices, and code-level decisions against the project artifacts, source files, experiment logs, and compiled manuscript. We take responsibility for the final content of the work.

\subsection*{Ethics statement}

This work studies recommendation-control algorithms using public or reproducible interaction resources and simulator-style rollout protocols. It does not introduce new human-subject data. Because target-state control can be used to steer future consumption statistics, deployment would require user consent, platform-governance review, and monitoring for distributional or group-level harms. The experiments measure future-state reasoning under explicit utility constraints, not unrestricted behavioral steering.

\subsection*{Reproducibility statement}

The closed-loop KuaiSim protocol uses fixed seeds, prediction depth $P=1$, interaction horizon $H=10$, slate size 6, a fixed diversity-rerank utility anchor, a learned state-action dynamics selector, target-aggressive semantic controls, an RFSS-U support-frontier reference, and a state-only dynamics ablation under the same target grid. MovieLens-25M and KuaiRand-Pure use the frozen chronological splits described in the experiments. To make the logged and closed-loop evidence reproducible, the manuscript specifies the selection rule, target-state definitions, validity criteria, and paired-seed interpretation. Source code, launch commands, metric JSON/CSV files, and figure-generation artifacts will be released as supplementary material.

\subsection{Logged Decision Times and Evaluation Windows}
\label{app:logged-temporal-protocol}

Each user's ordered record is divided into training, calibration, and test blocks. The calibration decision occurs at the end of the training block: training positives form the context and target, and calibration positives supply future outcomes. The test decision occurs at the end of the calibration block: training and calibration positives form the context and target, and test positives supply future outcomes. Thus, the calibration block has two roles at different decision times: future labels for calibration and observed history for test. The model input is clipped to the last 100 positive events; target construction uses the available positive history, and seen-item masking uses all interactions in the completed blocks.

For test selection, each training-positive event contributes one count to every associated genre or tag, and each calibration-positive event contributes $\omega=2$ counts. Normalizing the summed counts gives Equation~\ref{eq:prefix-target}. This is a block-level recency weighting: the more recent observed block receives twice the per-event weight. The weight is constant within each block; Appendix~\ref{app:logged-sensitivity} compares the tested history weights. Calibration scoring omits the calibration block from the target. Repeated positive events contribute repeatedly to the profile; an item with multiple tags contributes to each tag before normalization. A zero total count returns a zero profile.

The logged evaluation window is the entire subsequent held-out block. At test time, $\mathcal{F}^{+}_{u,t}$ is the list of all positive events in that user's test block. Its length varies by user; there is no additional fixed event-count or day-count cutoff. The future profile aggregates the tag counts of this list and normalizes them in the same way as the history profile. This window is distinct from the encoder history limit of 100, the slate size $K=20$, and the simulator interaction horizon $H=10$. Calibration analogously evaluates the complete calibration-positive block.

For window-sensitivity comparisons, the decision prefix and selected slate must remain fixed while the evaluation suffix changes. Fixed-length positive-event windows require a common eligible-user set across lengths and a recomputed full-block reference on that same set. This separates a change in evaluation horizon from a change in the evaluated population.

\subsection*{Algorithmic contract}

\begin{center}
  \fbox{\begin{minipage}{0.92\linewidth}
  \small
  \textbf{Algorithm 1: Utility-anchored target world model.}
  Given a state or prefix $h$, target $z$, frozen ranker $b$, and item statistics $G$:
  \begin{enumerate}
    \item Anchor: compute $A^0=\operatorname{TopK}_{i} f_b(h,i)$.
    \item Propose: build $\mathcal{C}(h,z)$ from $A^0$, anchor-local edits, and target-conditioned replacements.
    \item Predict: in replay, compute utility and target margins and estimate failure risk. In closed-loop control, predict one-step target components with $F_\theta(s,A)$ and obtain the immediate utility estimate.
    \item Select: retain candidates that pass the regime-specific selection conditions and choose the largest predicted target gain among them. If none qualifies, return $A^0$. A non-anchor selection issues a claim before outcome evaluation.
  \end{enumerate}
  \end{minipage}}
\end{center}

\begin{table}[!htbp]
  \caption{Implementation settings for logged replay and closed-loop rollout. Thresholds are selected on calibration users or calibration rollouts and evaluated once on held-out users or seeds.}
  \label{tab:implementation-settings}
  \centering
  \vspace{1mm}
  \small
  \setlength{\tabcolsep}{4.5pt}
  \renewcommand{\arraystretch}{1.05}
  \begin{tabular}{@{}lp{0.69\linewidth}@{}}
    \toprule
    Component & Setting \\
    \midrule
    Logged replay & Chronological split; maximum history length 100; slate size $K=20$; backbone top-$M$ candidate pool with $M=200$; candidate weights $\beta\in\{0.05,0.1,0.2,0.3,0.5,0.75,1.0\}$. \\
    Prefix target & $z^\dagger$ is computed from prefix-positive item statistics only. Calibration scoring uses train-prefix positives; test scoring uses train and calibration positives with calibration weight $\omega=2$. Held-out future positives are evaluation labels only. \\
    Logged gates & Utility-gain grid $\{-0.005,-0.002,-0.001,-0.0005,0,0.0002,0.0005\}$; state-gain grid $\{0,0.001,0.002,0.003,0.005,0.008\}$; risk grid $\{0.15,0.20,0.25,0.30,0.35,0.40,0.50,0.60\}$. \\
    Risk model & $L_2$-regularized logistic unsafe-claim classifier; learning rate 0.08; $L_2$ penalty 0.001; 300 optimization steps; features are $\beta$, utility margin, state margin, target concentration, interactions, negative utility margin, and squared margins. \\
    Split-risk diagnostic & Two classifiers estimate utility-floor violation and target-shortfall risk; the selector gates by the maximum predicted risk. \\
    KuaiSim dynamics model & Ridge state-action dynamics model trained on offline transition tuples before target-grid evaluation; features contain aggregate user/exposure state and slate-action summaries; targets contain feedback and next-state statistics. \\
    KuaiSim rollout & Seven target-state scenarios; ten seeds; prediction depth $P=1$; interaction horizon $H=10$; slate size $K=6$; diversity-rerank utility anchor; nine candidate actions per decision: the anchor, two anchor-local edits, and six graph-semantic replacement slates. \\
    Graph-semantic proposals & Fixed proposal families over observable click scores, user-history repeat indicators, item-tag overlap with the user history, and within-slate item-embedding diversity. Proposal families instantiate repeat, balanced, and diversity directions under preset click-floor variants; rollout outcomes are not used to construct the candidates. \\
    Closed-loop controls & The reported same-protocol controls are the fixed diversity-rerank anchor, two target-aggressive semantic policies, an RFSS-U support-frontier controller, and a state-only learned-dynamics ablation. \method{} uses the learned dynamics model to select a candidate slate; KuaiSim executes the selected slate and provides realized feedback and next-state metrics for evaluation. \\
    \bottomrule
  \end{tabular}
\end{table}

\begin{table}[!htbp]
  \caption{Seven-scenario KuaiSim target grid for closed-loop evaluation. Each target is a desired value for repeat exposure, tag entropy, history-tag overlap, and cumulative coverage; the status column records whether the fixed support-frontier reference finds a target-control action under the same validity test. Repeat is the fraction of slate items already seen in the user history. Entropy and history overlap are computed from item tags. Coverage is the cumulative number of distinct exposed items over the rollout.}
  \label{tab:kuaisim-target-grid}
  \centering
  \vspace{1mm}
  \footnotesize
  \setlength{\tabcolsep}{3.8pt}
  \renewcommand{\arraystretch}{1.04}
  \begin{tabular}{@{}lrrrrl@{}}
    \toprule
    Scenario & Repeat & Entropy & Hist. overlap & Coverage & Status \\
    \midrule
    balanced\_anchor & 0.0400 & 3.300 & 0.580 & 650 & partial \\
    diverse\_anchor & 0.0325 & 3.425 & 0.530 & 735 & feasible \\
    high\_div\_frontier & 0.0250 & 3.460 & 0.510 & 780 & feasible \\
    midpoint\_div\_bal & 0.0363 & 3.360 & 0.555 & 693 & feasible \\
    mild\_repeat\_frontier & 0.0500 & 3.320 & 0.575 & 630 & feasible \\
    partial\_low\_entropy & 0.0450 & 3.100 & 0.620 & 600 & unsupported \\
    repeat\_unreachable & 0.0750 & 2.550 & 0.720 & 550 & unsupported \\
    \bottomrule
  \end{tabular}
\end{table}

\FloatBarrier
\section{Evidence Map}
\label{app:evidence-map}

The evaluation separates cross-backbone transfer, selection behavior, and interactive control. Logged replay measures ranking and future-statistic alignment on fixed chronological histories. Same-support diagnostics isolate the effects of the selector's components. KuaiSim evaluates repeated decisions with action-dependent feedback and state updates.

\medskip \noindent\textbf{Logged consequence transfer.} Table~\ref{tab:main-logged-results} pairs each backbone with its consequence-aware version on MovieLens-25M and KuaiRand-Pure. All twelve pairs improve Recall@20, NDCG@20, and future-state alignment in both domains. The fixed-backbone comparison measures the contribution of the decision layer across model families.

\medskip \noindent\textbf{Constrained consequence selection.} Coverage measures how often the selector leaves the anchor; replay-unsafe rate measures failure among those selections. Jointly reporting them distinguishes reliable target-directed choices from low failure rates obtained through frequent fallback.

\medskip \noindent\textbf{Closed-loop claim validity.} KuaiSim evaluates whether an issued target-control claim satisfies the utility and target conditions after interaction. The target grid includes different requested directions and support conditions, allowing both intervention and fallback to be evaluated.

The two regimes use different outcome sources: historical future positives in replay and feedback from executed actions in rollout. Their results characterize transfer and simulator-based control, respectively.

\begin{table}[!htbp]
  \caption{Action-conditioned dynamics improve KuaiSim transition prediction. Normalized MAE is averaged over feedback and next-state targets; lower is better.}
  \label{tab:dynamics-prediction}
  \centering
  \vspace{1mm}
  \small
  \setlength{\tabcolsep}{4.4pt}
  \renewcommand{\arraystretch}{1.03}
  \begin{tabular}{@{}llrrr@{}}
    \toprule
    Split & Model & State only & State + action & Relative reduction \\
    \midrule
    held-out seed & ridge & 0.491 & \textbf{0.266} & 45.8\% \\
    held-out seed & random forest & 0.307 & \textbf{0.287} & 6.6\% \\
    held-out policy & ridge & 0.433 & \textbf{0.345} & 20.4\% \\
    held-out policy & random forest & \textbf{0.468} & 0.486 & -3.9\% \\
    \bottomrule
  \end{tabular}
\end{table}

Including action summaries reduces ridge prediction error on both held-out seeds and held-out policies. Random forests improve on held-out seeds but worsen on held-out policies, so the benefit of action conditioning depends on the predictor and evaluation split. The target-grid controller uses ridge dynamics. Prediction error and executed control outcomes evaluate complementary parts of this pipeline.

\clearpage
\section{Additional Logged Analyses}
\label{app:additional-logged-analyses}

\begin{figure}[!htbp]
  \centering
  \includegraphics[width=\linewidth]{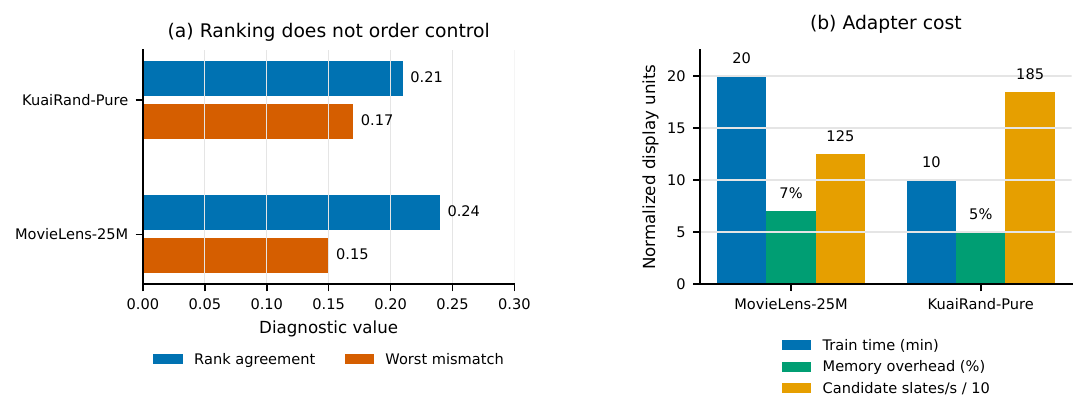}
  \caption{Additional logged replay diagnostics: (a) ranking--state ordering on the base backbones and (b) adapter training time, search throughput, and memory overhead.}
  \label{fig:appendix-logged-diagnostics}
\end{figure}

The additional logged analyses examine ranking--state ordering, aggressive versus constrained selection, parameter sensitivity, and computational cost. Table~\ref{tab:same-support-ablation} reports matched policy contrasts, Table~\ref{tab:wm-quality} evaluates risk prediction, and Figure~\ref{fig:appendix-logged-diagnostics} summarizes the remaining diagnostics.

\begin{table}[H]
  \caption{Same-support ablation of aggressive and constrained world-model selection. Constrained selection retains positive ranking and state gains while reducing unsafe target-control claims across both logged domains. R@20 and N@20 denote Recall@20 and NDCG@20. $\Delta S$ is the reduction in future-state $L_1$ loss, so larger is better. Bold marks the better value within each aggressive/constrained pair.}
  \label{tab:same-support-ablation}
  \centering
  \vspace{1mm}
  \tiny
  \setlength{\tabcolsep}{2.8pt}
  \renewcommand{\arraystretch}{1.02}
  \resizebox{\textwidth}{!}{%
  \begin{tabular}{@{}llrrrrrrrr@{}}
    \toprule
    Backbone & Policy & $\Delta$R@20 & 95\% CI & $\Delta$N@20 & 95\% CI & $\Delta S$ & 95\% CI & Cov. & Inv. \\
    \midrule
    \multicolumn{10}{@{}l}{\textit{MovieLens-25M}} \\
    GRU4Rec & Aggressive WM & 0.0119 & [0.0114, 0.0124] & 0.0034 & [0.0032, 0.0036] & \textbf{0.045} & [0.043, 0.047] & \textbf{83.5} & 14.2 \\
     & Constrained WM & \textbf{0.0127} & [0.0122, 0.0132] & \textbf{0.0037} & [0.0035, 0.0039] & 0.041 & [0.039, 0.043] & 56.2 & \textbf{3.4} \\
    SASRec & Aggressive WM & 0.0114 & [0.0109, 0.0119] & 0.0040 & [0.0038, 0.0042] & \textbf{0.045} & [0.043, 0.047] & \textbf{82.0} & 15.1 \\
     & Constrained WM & \textbf{0.0127} & [0.0123, 0.0131] & \textbf{0.0044} & [0.0042, 0.0046] & 0.041 & [0.039, 0.043] & 54.7 & \textbf{3.6} \\
    BERT4Rec & Aggressive WM & 0.0111 & [0.0107, 0.0115] & 0.0045 & [0.0043, 0.0047] & \textbf{0.044} & [0.042, 0.046] & \textbf{95.0} & 8.5 \\
     & Constrained WM & \textbf{0.0127} & [0.0122, 0.0132] & \textbf{0.0051} & [0.0049, 0.0053] & 0.041 & [0.039, 0.043] & 94.2 & \textbf{5.1} \\
    NextItNet & Aggressive WM & 0.0112 & [0.0108, 0.0116] & 0.0037 & [0.0035, 0.0039] & \textbf{0.045} & [0.043, 0.047] & \textbf{84.0} & 13.7 \\
     & Constrained WM & \textbf{0.0127} & [0.0122, 0.0132] & \textbf{0.0041} & [0.0039, 0.0043] & 0.041 & [0.039, 0.043] & 55.8 & \textbf{3.5} \\
    FPMC & Aggressive WM & 0.0134 & [0.0129, 0.0139] & 0.0032 & [0.0030, 0.0034] & \textbf{0.051} & [0.049, 0.053] & \textbf{86.0} & 12.5 \\
     & Constrained WM & \textbf{0.0144} & [0.0139, 0.0149] & \textbf{0.0035} & [0.0033, 0.0037] & 0.046 & [0.044, 0.048] & 57.8 & \textbf{3.5} \\
    TransRec & Aggressive WM & 0.0120 & [0.0115, 0.0125] & 0.0026 & [0.0024, 0.0028] & \textbf{0.047} & [0.045, 0.049] & \textbf{83.0} & 14.0 \\
     & Constrained WM & \textbf{0.0130} & [0.0125, 0.0135] & \textbf{0.0028} & [0.0026, 0.0030] & 0.042 & [0.040, 0.044] & 53.1 & \textbf{3.8} \\
    STAMP & Aggressive WM & 0.0115 & [0.0110, 0.0120] & 0.0029 & [0.0027, 0.0031] & \textbf{0.042} & [0.040, 0.044] & \textbf{84.5} & 13.0 \\
     & Constrained WM & \textbf{0.0125} & [0.0120, 0.0130] & \textbf{0.0031} & [0.0029, 0.0033] & 0.037 & [0.035, 0.039] & 52.4 & \textbf{4.1} \\
    LightSANs & Aggressive WM & 0.0109 & [0.0105, 0.0113] & 0.0042 & [0.0040, 0.0044] & \textbf{0.045} & [0.043, 0.047] & \textbf{95.5} & 8.0 \\
     & Constrained WM & \textbf{0.0119} & [0.0114, 0.0124] & \textbf{0.0047} & [0.0045, 0.0049] & 0.040 & [0.038, 0.042] & 93.8 & \textbf{5.2} \\
    S3Rec & Aggressive WM & 0.0110 & [0.0106, 0.0114] & 0.0042 & [0.0040, 0.0044] & \textbf{0.046} & [0.044, 0.048] & \textbf{95.0} & 8.2 \\
     & Constrained WM & \textbf{0.0120} & [0.0115, 0.0125] & \textbf{0.0046} & [0.0044, 0.0048] & 0.041 & [0.039, 0.043] & 94.4 & \textbf{5.1} \\
    SINE & Aggressive WM & 0.0111 & [0.0107, 0.0115] & 0.0029 & [0.0027, 0.0031] & \textbf{0.043} & [0.041, 0.045] & \textbf{88.0} & 11.0 \\
     & Constrained WM & \textbf{0.0121} & [0.0116, 0.0126] & \textbf{0.0031} & [0.0029, 0.0033] & 0.038 & [0.036, 0.040] & 63.5 & \textbf{3.3} \\
    ADRec & Aggressive WM & 0.0115 & [0.0110, 0.0120] & 0.0084 & [0.0080, 0.0088] & \textbf{0.046} & [0.044, 0.048] & \textbf{75.0} & 16.0 \\
     & Constrained WM & \textbf{0.0127} & [0.0122, 0.0132] & \textbf{0.0099} & [0.0095, 0.0103] & 0.041 & [0.039, 0.043] & 41.5 & \textbf{4.2} \\
    FAVE & Aggressive WM & 0.0115 & [0.0110, 0.0120] & 0.0092 & [0.0088, 0.0096] & \textbf{0.045} & [0.043, 0.047] & \textbf{95.0} & 8.5 \\
     & Constrained WM & \textbf{0.0127} & [0.0122, 0.0132] & \textbf{0.0102} & [0.0098, 0.0106] & 0.041 & [0.039, 0.043] & 94.0 & \textbf{5.0} \\
    \addlinespace[0.6mm]
    \multicolumn{10}{@{}l}{\textit{KuaiRand-Pure}} \\
    GRU4Rec & Aggressive WM & 0.0052 & [0.0038, 0.0066] & 0.0020 & [0.0014, 0.0026] & \textbf{0.078} & [0.072, 0.084] & \textbf{96.5} & 17.5 \\
     & Constrained WM & \textbf{0.0058} & [0.0044, 0.0072] & \textbf{0.0024} & [0.0018, 0.0030] & 0.072 & [0.066, 0.078] & 95.7 & \textbf{5.1} \\
    SASRec & Aggressive WM & 0.0051 & [0.0037, 0.0065] & 0.0019 & [0.0013, 0.0025] & \textbf{0.077} & [0.071, 0.083] & \textbf{96.8} & 17.0 \\
     & Constrained WM & \textbf{0.0058} & [0.0044, 0.0072] & \textbf{0.0023} & [0.0017, 0.0029] & 0.072 & [0.066, 0.078] & 96.0 & \textbf{5.2} \\
    BERT4Rec & Aggressive WM & 0.0051 & [0.0037, 0.0065] & 0.0020 & [0.0014, 0.0026] & \textbf{0.077} & [0.071, 0.083] & \textbf{98.5} & 16.0 \\
     & Constrained WM & \textbf{0.0058} & [0.0045, 0.0071] & \textbf{0.0021} & [0.0015, 0.0027] & 0.072 & [0.067, 0.077] & 98.2 & \textbf{6.0} \\
    NextItNet & Aggressive WM & 0.0050 & [0.0036, 0.0064] & 0.0020 & [0.0014, 0.0026] & \textbf{0.077} & [0.071, 0.083] & \textbf{96.2} & 17.8 \\
     & Constrained WM & \textbf{0.0058} & [0.0043, 0.0073] & \textbf{0.0023} & [0.0017, 0.0029] & 0.072 & [0.066, 0.078] & 95.5 & \textbf{5.0} \\
    FPMC & Aggressive WM & 0.0058 & [0.0044, 0.0072] & 0.0021 & [0.0015, 0.0027] & \textbf{0.084} & [0.078, 0.090] & 98.0 & 15.0 \\
     & Constrained WM & \textbf{0.0068} & [0.0054, 0.0082] & \textbf{0.0023} & [0.0017, 0.0029] & 0.079 & [0.073, 0.085] & \textbf{98.9} & \textbf{5.0} \\
    TransRec & Aggressive WM & 0.0048 & [0.0034, 0.0062] & 0.0017 & [0.0011, 0.0023] & \textbf{0.087} & [0.081, 0.093] & \textbf{96.8} & 17.2 \\
     & Constrained WM & \textbf{0.0058} & [0.0044, 0.0072] & \textbf{0.0020} & [0.0014, 0.0026] & 0.082 & [0.076, 0.088] & 95.2 & \textbf{5.5} \\
    STAMP & Aggressive WM & 0.0047 & [0.0033, 0.0061] & 0.0018 & [0.0012, 0.0024] & \textbf{0.069} & [0.063, 0.075] & \textbf{97.0} & 16.5 \\
     & Constrained WM & \textbf{0.0057} & [0.0043, 0.0071] & \textbf{0.0020} & [0.0014, 0.0026] & 0.064 & [0.058, 0.070] & 96.1 & \textbf{5.9} \\
    LightSANs & Aggressive WM & 0.0045 & [0.0031, 0.0059] & 0.0016 & [0.0010, 0.0022] & \textbf{0.079} & [0.073, 0.085] & \textbf{98.0} & 16.8 \\
     & Constrained WM & \textbf{0.0055} & [0.0041, 0.0069] & \textbf{0.0019} & [0.0013, 0.0025] & 0.074 & [0.068, 0.080] & 97.4 & \textbf{6.1} \\
    S3Rec & Aggressive WM & 0.0044 & [0.0030, 0.0058] & 0.0016 & [0.0010, 0.0022] & \textbf{0.069} & [0.063, 0.075] & \textbf{99.0} & 15.5 \\
     & Constrained WM & \textbf{0.0054} & [0.0040, 0.0068] & \textbf{0.0018} & [0.0012, 0.0024] & 0.064 & [0.058, 0.070] & 98.6 & \textbf{6.0} \\
    SINE & Aggressive WM & 0.0045 & [0.0031, 0.0059] & 0.0017 & [0.0011, 0.0023] & \textbf{0.073} & [0.067, 0.079] & \textbf{99.2} & 15.8 \\
     & Constrained WM & \textbf{0.0055} & [0.0041, 0.0069] & \textbf{0.0019} & [0.0013, 0.0025] & 0.068 & [0.062, 0.074] & 99.0 & \textbf{5.6} \\
    ADRec & Aggressive WM & -0.0021 & [-0.0035, -0.0007] & -0.0018 & [-0.0025, -0.0011] & \textbf{0.110} & [0.106, 0.114] & \textbf{98.5} & 10.5 \\
     & Constrained WM & \textbf{0.0058} & [0.0044, 0.0072] & \textbf{0.0036} & [0.0029, 0.0043] & 0.072 & [0.067, 0.077] & 94.8 & \textbf{5.8} \\
    FAVE & Aggressive WM & -0.0013 & [-0.0027, 0.0001] & -0.0011 & [-0.0018, -0.0004] & \textbf{0.113} & [0.109, 0.117] & \textbf{99.2} & 10.8 \\
     & Constrained WM & \textbf{0.0058} & [0.0043, 0.0073] & \textbf{0.0037} & [0.0030, 0.0044] & 0.072 & [0.066, 0.078] & 98.1 & \textbf{6.1} \\
    \bottomrule
  \end{tabular}
  }
\end{table}

The first diagnostic is a negative result about the base ranking objective. For each logged domain, we compare the ordering induced by next-item utility with the ordering induced by future-state alignment over the same candidate support. Low rank agreement means that a high-utility slate can still be a poor control action for the future statistic. This is the empirical reason for separating utility gain and state gain in the consequence layer instead of treating the backbone score as a sufficient action value.

Section~\ref{app:logged-sensitivity} reports pointwise sensitivity to candidate-pool size, proposal grids, history weighting, and evaluation-window length. Figure~\ref{fig:appendix-logged-diagnostics} reports ranking diagnostics and adapter training, search throughput, and memory overhead, separating the additional decision-layer cost from backbone training.

\begin{table}[!htbp]
  \caption{Risk models identify unsafe slate claims before selection. Split-risk gating lowers selected unsafe rates and improves unsafe-candidate ranking in the representative logged settings. Unsafe and predicted risk are percentages. AUC measures unsafe-candidate ranking; lower ECE indicates better calibration.}
  \label{tab:wm-quality}
  \centering
  \vspace{1mm}
  \small
  \setlength{\tabcolsep}{0pt}
  \renewcommand{\arraystretch}{1.02}
  \begin{tabular*}{\linewidth}{@{\extracolsep{\fill}}llllrrrr@{}}
    \toprule
    Dataset & Backbone & Risk model & Scope & Unsafe & Pred. risk & AUC & ECE \\
    \midrule
    MovieLens-25M & S3Rec & single-risk & all changed & 22.6 & 33.3 & 0.739 & 0.107 \\
     &  & single-risk & selected & 5.9 & 19.4 & 0.811 & 0.135 \\
     &  & split-risk & selected & \textbf{4.1} & 13.8 & \textbf{0.825} & \textbf{0.099} \\
    \addlinespace[0.4mm]
    KuaiRand-Pure & LightSANs & single-risk & all changed & 7.7 & 21.8 & 0.730 & 0.141 \\
     &  & single-risk & selected & 6.4 & 16.2 & 0.743 & 0.098 \\
     &  & split-risk & selected & \textbf{4.8} & 14.0 & \textbf{0.802} & \textbf{0.097} \\
    \bottomrule
  \end{tabular*}
\end{table}

Together, these diagnostics characterize the logged selector through its ranking gains, state alignment, risk estimates, and selection frequency. Their common candidate support allows changes in these quantities to be attributed to different decision rules within the replay protocol.

\FloatBarrier
\section{Closest-Work Distinctions}
\label{app:closest-work}

\method{} is closest to work on interactive recommendation, model-based recommendation, goal-conditioned recommendation, causal feedback correction, and LLM-based user simulation. These lines overlap with parts of the problem, but they optimize or evaluate different objects.

Interactive simulators such as KuaiSim and RecSim provide environments in which slate actions can produce future feedback. UA-TWM uses this interaction to evaluate a local controller around a trained recommendation anchor. Model-based online recommendation learns user-agent interaction models for policy optimization \citep{bai2019modelbasedrec}. SlateQ-style and DORL-style methods optimize long-term or penalized return. In contrast, \method{} starts from a strong sequential ranking anchor and asks whether nearby slate actions can be selected by predicted utility gain, future-state gain, and replay-unsafe risk.

Goal-conditioned recommendation is also related. GoalRec learns goal-conditioned long-term value estimates \citep{wang2021goalrec}, while more recent goal-conditioned sequence models condition action prediction on desired cumulative objectives \citep{li2024goalconditionedrec,gao2025mocdt}. UA-TWM evaluates target-directed selection within the anchor's candidate support, using an explicit utility floor. Causal and feedback-loop methods such as CIRS correct bias or satisfaction dynamics under intervention. LLM-as-judge and AlignUSER-style systems use language models to simulate or evaluate users. \method{} instead studies a non-LLM consequence layer whose output is constrained by utility preservation, support risk, and fallback.

\section{Closed-Loop Mechanism Details}
\label{app:closed-loop-details}
\label{app:mechanism-ablation-figure}
\label{app:decision-trace}

Closed-loop evaluation records target-scenario claims separately from individual slate choices. Selecting a non-anchor policy issues a claim before its rollout outcome is observed; Equation~\ref{eq:claim-event} subsequently determines validity. Coverage counts all issued claims, including unsuccessful ones, and invalidity is the fraction that fails the utility or target condition. Anchor fallback issues no claim. Decision-level traces instead count candidate sources selected at individual simulator steps.

Table~\ref{tab:closed-loop-diagnostic} exhibits distinct utility--target trade-offs. The target-aggressive policies achieve lower target error at lower click utility. The state-only ablation retains the anchor on all scenarios. The full controller makes four valid target-scenario claims and returns the anchor on three scenarios. Proposal generation supplies the available alternatives, while action-conditioned scoring and the utility check determine which alternatives are selected.

\begin{table}[!htbp]
  \caption{Hidden target-family audit for closed-loop target control. \method{} claims utility-safe movement on feasible target families and falls back on the unsupported entropy target. Click gap is selected click minus anchor click. Control gain is anchor target error minus selected target error.}
  \label{tab:hidden-target-family}
  \centering
  \vspace{1mm}
  \small
  \setlength{\tabcolsep}{4.2pt}
  \renewcommand{\arraystretch}{1.02}
  \begin{tabular}{@{}lrrrrrr@{}}
    \toprule
    Target family & Scenarios & Claims & Invalid & Selected & Click gap & Control gain \\
    \midrule
    balanced & 1 & 1 & 0 & 1 & 0.111 & 0.403 \\
    diversity & 2 & 1 & 0 & 1 & 0.022 & 0.236 \\
    entropy & 1 & 0 & 0 & 0 & 0.000 & 0.000 \\
    mixed diversity-balance & 1 & 1 & 0 & 1 & 0.047 & 0.476 \\
    repeat & 2 & 1 & 0 & 1 & 0.052 & 0.158 \\
    \bottomrule
  \end{tabular}
\end{table}

Table~\ref{tab:hidden-target-family} resolves the aggregate claim count by requested target family. The controller makes a valid claim in the balanced, diversity, mixed, and repeat families, and falls back on the entropy target. Fallback marks a target for which the controller selects no qualifying alternative within the evaluated local support; global reachability remains undetermined.

\begin{figure}[!htbp]
  \centering
  \includegraphics[width=\linewidth]{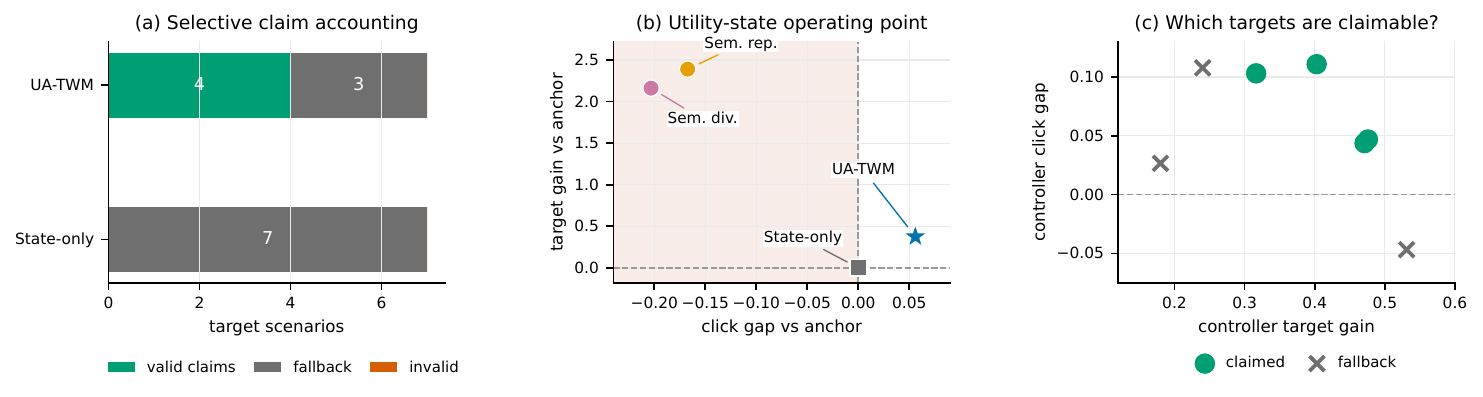}
  \caption{Closed-loop mechanism diagnostic from the leave-evaluation-seed learned-dynamics run. Panel (a) separates target-scenario claims from fallback decisions. Panel (b) compares target-aggressive controls, the state-only dynamics ablation, and \method{} in the utility-state plane. Panel (c) shows the seven hidden target scenarios by realized controller gain and click gap.}
  \label{fig:mechanism-ablation}
\end{figure}

\begin{figure}[!htbp]
  \centering
  \includegraphics[width=0.62\linewidth]{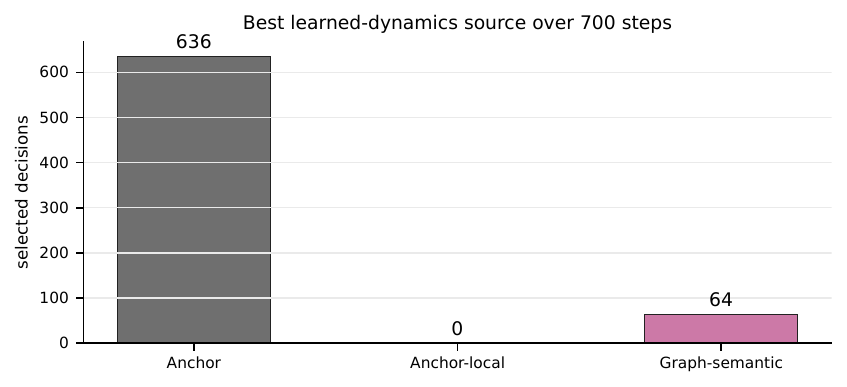}
  \caption{Decision-level source audit for the same closed-loop run. Counts are unique by target scenario, seed, and simulator step. The controller evaluates 6,300 candidate actions over 700 decisions and selects only actions from the declared support.}
  \label{fig:decision-trace}
\end{figure}

Figure~\ref{fig:mechanism-ablation} relates scenario-level target gains to utility and claim outcomes. Figure~\ref{fig:decision-trace} records the sources of executed actions within the declared candidate support. The two views connect aggregate controller performance with its individual decisions.

\section{Logged Sensitivity Analysis}
\label{app:logged-sensitivity}

\begin{figure}[!htbp]
  \centering
  \includegraphics[width=\linewidth]{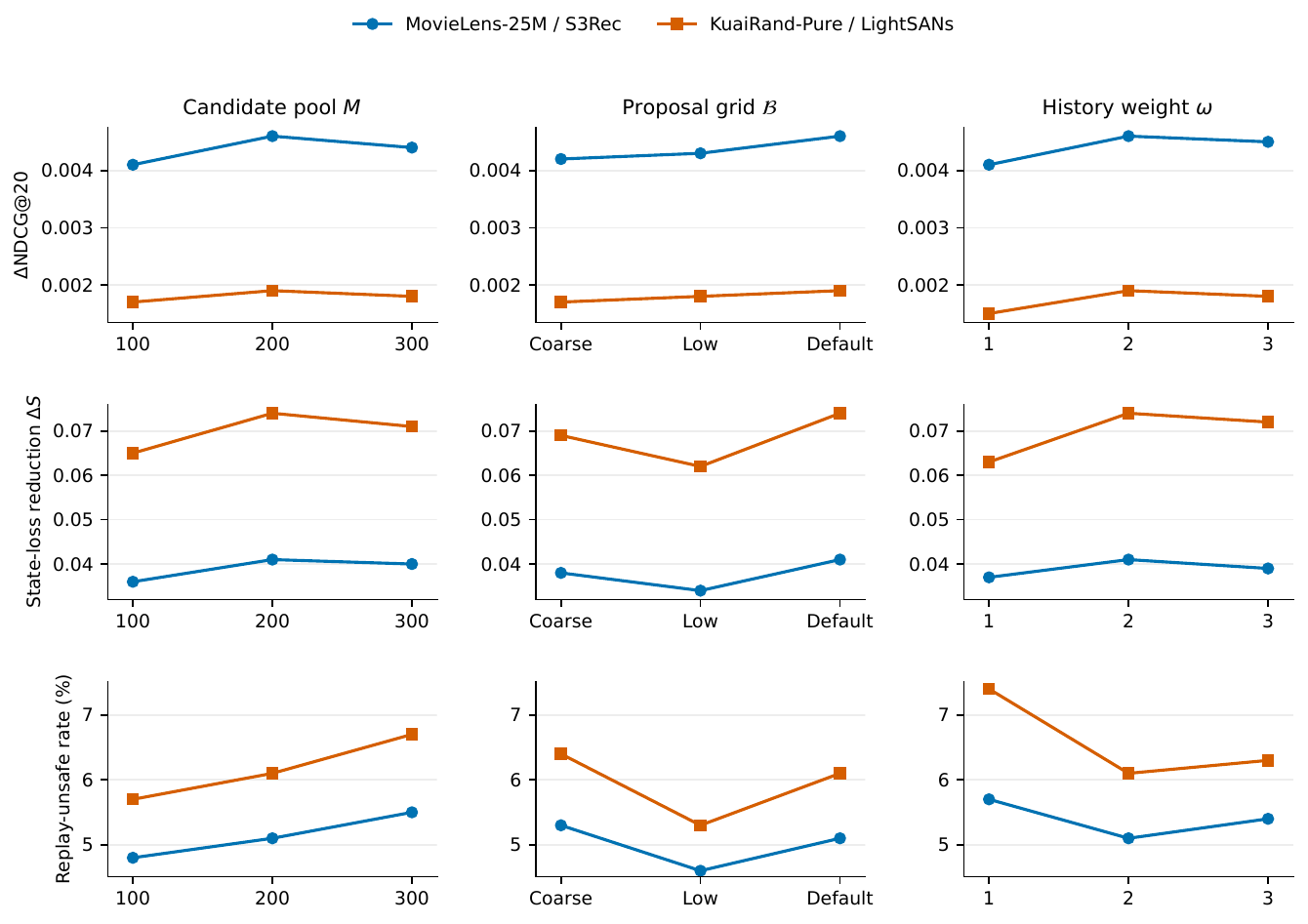}
  \caption{Sensitivity to candidate-pool size, proposal grid, and history weight. Rows show NDCG gain, state-loss reduction, and replay-unsafe rate for two dataset--backbone settings. Gains are relative to the corresponding frozen backbone.}
  \label{fig:logged-sensitivity}
\end{figure}

We vary one factor at a time on MovieLens-25M/S3Rec and KuaiRand-Pure/LightSANs. The backbone remains fixed; the adapter and selection thresholds are recalibrated using calibration data for each candidate-pool, proposal-grid, or history-weight configuration. Default values are $M=200$, $\omega=2$, and $\mathcal{B}=\{0.05,0.1,0.2,0.3,0.5,0.75,1.0\}$. The coarse grid is $\{0.1,0.3,1.0\}$, and the low-strength grid is $\{0.05,0.1,0.2,0.3,0.5\}$. Figure~\ref{fig:logged-sensitivity} presents measured operating points; the grid labels denote discrete configurations.

\begin{samepage}
\paragraph{Candidate support and history weighting.}
Across the seven configurations per setting, MovieLens Recall gains range from 0.0107 to 0.0120, NDCG gains from 0.0041 to 0.0046, and state-loss reductions from 0.034 to 0.041. KuaiRand gains range from 0.0046 to 0.0055, 0.0015 to 0.0019, and 0.062 to 0.074, respectively. Expanding the pool from 200 to 300 does not improve these gains and increases unsafe rates from 5.1\% to 5.5\% and from 6.1\% to 6.7\%. The larger pool adds candidates without improving the selected operating point in these settings. The low-strength grid lowers unsafe rates to 4.6\% and 5.3\%, while reducing state gains to 0.034 and 0.062. History weights of one and three preserve positive gains, with weight two giving the largest observed state reduction in this sweep. The sweep characterizes sensitivity within these tested ranges.

\end{samepage}

\paragraph{Evaluation-window length.}
For each dataset, the window experiment uses the same users with at least 20 test-positive events. The policy, decision prefix, and selected slate are fixed; evaluation uses the first 10 events, the first 20 events, or the full test-positive block. The full-block reference is recomputed on this common cohort for comparison with the shorter windows. Coverage remains 95.7\% on MovieLens and 98.2\% on KuaiRand across all three windows, as expected when only outcomes change.

\begin{figure}[!htbp]
  \centering
  \includegraphics[width=\linewidth]{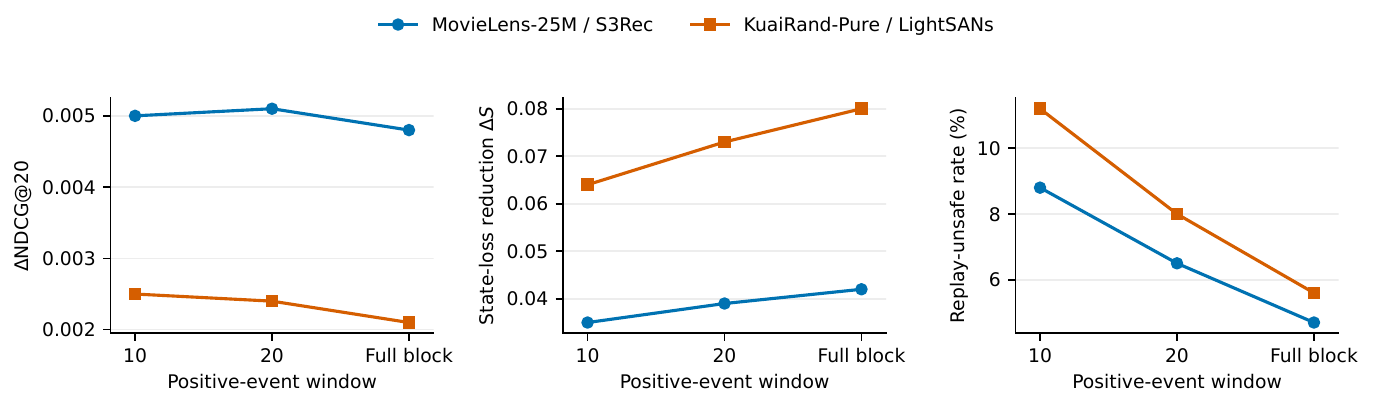}
  \caption{Evaluation-window sensitivity with fixed policies and matched users. NDCG and state gains remain positive; shorter windows yield higher replay-unsafe rates. Full block denotes each user's complete test-positive suffix.}
  \label{fig:logged-window-sensitivity}
\end{figure}

MovieLens state-loss reduction increases from 0.035 at 10 events to 0.039 at 20 events and 0.042 over the full block; unsafe rates fall from 8.8\% to 6.5\% and 4.7\%. KuaiRand shows the same direction, with state reductions of 0.064, 0.073, and 0.080 and unsafe rates of 11.2\%, 8.0\%, and 5.6\%. Ranking gains remain positive in every window, although NDCG gains are not monotonic in window length. The invariant coverage and changing invalidity separate selection frequency from the timescale over which the selected slate is evaluated.

\end{document}